\documentclass[%
 aip,
amsmath,amssymb,
preprint,%
]{revtex4-1}

\usepackage{graphicx}
\usepackage{dcolumn}
\usepackage{bm}

\usepackage[utf8]{inputenc}
\usepackage[T1]{fontenc}

\usepackage{graphicx}
\usepackage{amsfonts,mathrsfs,amsmath,amssymb,bbold,mathtools}
\usepackage{bm}
\usepackage{tabulary}
\usepackage{color}
\usepackage{graphicx}
\usepackage{xfrac}
\usepackage{pifont}
\usepackage{ulem}

\newif\ifK
\Ktrue

\begin{document}

\preprint{AIP/123-QED}

\title[]{Multiparameter quantum metrology with postselection measurements}

\author{Le Bin Ho}
\affiliation{ 
		Department of Physics, Kindai University, 
		Higashi-Osaka, 577-8502, Japan.}
\affiliation{Research Institute of Electrical Communication, Tohoku University, Sendai, 980-8577, Japan.}
\email{binho@riec.tohoku.ac.jp} 
              
\author{Yasushi Kondo}%
\affiliation{ 
	      Department of Physics, Kindai University, 
              Higashi-Osaka, 577-8502, Japan.}
\affiliation{               
              Interdisciplinary Graduate School of Science and Engineering, 
              Kindai University, Higashi-Osaka, 577-8502, Japan
}%

\date{\today}

\begin{abstract}
We analyze simultaneous quantum estimations of multiple parameters
with postselection measurements in terms of a tradeoff relation.  
The system, or a sensor, is characterized by a set of parameters,  
interacts with a measurement apparatus (MA), and then
is postselected onto a set of orthonormal final states.
Measurements of the MA yield an estimation of the parameters.
We first derive classical and quantum Cram\'er-Rao lower bounds
and then discuss their archivable condition and the tradeoffs in the postselection measurements in general,
including the case when a sensor is in mixed state. 
Its whole information can, in principle, be obtained 
via the MA which is not possible without postselection. 
We, then, apply the framework to simultaneous measurements of
phase and its fluctuation as an example. 
\end{abstract}


\maketitle

\section {Introduction}\label{seci}
Quantum metrology is a promising technology.   
It is applicable for a wide range of fields, such as 
quantum sensing \cite{Degen89,Pezze90}, 
quantum imaging \cite{Preza16}, and 
detecting gravitational waves \cite{Aasi7,Branford121,Qui96, Miao9}.
The estimation of a single parameter has already been established 
\cite{Pezze102,Huelga79,Wineland46,Wineland50,
Giovannetti306,Giovannetti96,Jones324,Simmons82,
Zaiser7,Matsuzaki120}. 
Therein, several studies demonstrated the quantum-enhanced metrology 
by using entangled resources \cite{Pezze102,Huelga79,Wineland46,Wineland50,
Giovannetti306,Giovannetti96,Jones324,Simmons82}, 
quantum memory \cite{Zaiser7}, or teleportation \cite{Matsuzaki120}. 
Note, however, that it is often demanded simultaneous multiparameter estimations 
in many practical applications.
For example, estimation of phases 
are always affected by environmental noise  
and thus, a simultaneous measurement of the phase and 
its fluctuation is necessary. 
Such joint estimations have been discussed recently
\cite{Vidrighin5,Altorio92,Szczykulska2,Sergey2013,Crowley89,
Roccia2018,Pinel88,Gagatsos96}. 
Also, the current generation of laser-interferometric gravitational wave detectors,
such as LIGO \cite{Abadie7, Aasi7}, 
are inevitably affected by squeezing noises and optical loss, 
therefore the improvement of the detectors can be expected 
when the phase and its loss are measured simultaneously.
Furthermore, various practical applications have been discussed, 
including damping and temperature \cite{Monras83}, 
two-phase spin rotation \cite{Vaneph1}, 
waveform \cite{Berry5}, 
operators \cite{Fujiwara65,Ballester69}, 
phase-space displacements \cite{Genoni87,Steinlechner7}.
The estimations of multiple phases \cite{Humphreys111,Liu2} and 
parameters in multidimensional fields \cite{Baumgratz116, Ho2020} have been
discussed, too. 

Typically, a sensor characterized by multiple parameters 
will be measured by a set of POVMs to estimate the parameters.
We term it as ``direct sensing'' because no ancillary systems are 
required. The limit of the estimation precision is imposed 
by quantum mechanics and bounded by a so-called
quantum Cram\'er-Rao bound (QCRB)
\cite{Holevo2011,Paris7}. 
For a single-parameter estimation, the QCRB can be achieved
by projecting the states of a sensor on the basis determined by
eigenvectors of its symmetric logarithmic derivative (SLD) operator 
\cite{Vidrighin5,Crowley89,Dominik97}. 
However, for a multiparameter estimation,   
SLD operators of different parameters may not commute and thus 
the basis determined by them may not be orthonormal.
Such a case leads to a tradeoff in
the estimation of different parameters
\cite{Crowley89,Szczykulska2,Vidrighin5,Altorio92}. 
This tradeoff is a kind of ``competition'' among them 
\cite{Szczykulska1}. 
Although several theoretical and numerical studies of 
the optimal POVMs that saturate the QCRB 
as well as the tradeoff relations
have been reported \cite{Szczykulska2,Yang2018,Pezze119}, 
achieving the QCRB is still a challenging task in the multiparameter 
estimations. 
Alternatively, one can consider a so-called 
Holevo Cram\'er-Rao bound, which is asymptotically achievable 
and can reach twice times larger than the QCRB. 
See \cite{Rafal2020} and references therein.

In contrast to the ``direct sensing,'' an ``indirect one'' 
with postselection is also possible for parameter estimations,
hereafter referred to as a postselection measurement. 
The sensor interacts with an ancillary 
system, referred to as a measurement apparatus (MA). After the interaction, 
the sensor will be postselected while the final MA state will be measured 
to provide the estimation.
Postselection measurements 
have been used to estimate single parameters 
\cite{Knee87,Tanaka88,Knee4,Combes89,Ferrie112,Zhang114,Chen121,Pang113,
Pang92,Alves91,Alves95,Dressel88,Lyons114,Wang117,Pang115,Jordan4,Jordan2,
Pang94,Harris118,Viza92,Sinclair96,Ho383} and 
various methods have been proposed, 
including the optimal choices of the system and MA states \cite{Alves91,Alves95}, 
entangled sensors \cite{Pang113,Pang92}, 
photon recycling \cite{Dressel88,Lyons114,Wang117}, 
non-classical MA \cite{Pang115} to improve the precision.
There are, however, ongoing debates over the merit of postselection measurements
if it defeats the ultimate-limit precision or not. 
Most studies point out that postselection measurements 
cannot provide any advantages for the estimations
\cite{Knee87,Tanaka88,Knee4,Combes89,Ferrie112,Zhang114,
Chen121,Pang113,Pang92}. 
It is true because the postselection measurements do not 
generate ``information''. 
Thus, they alone cannot beat it, 
even though all the data (from the success and failure postselections) 
are taken into account \cite{Knee87,Combes89,Ferrie112,Zhang114,Chen121}. 
For example, Knee et al. \cite{Knee87,Knee4},
Tanaka and Yamamoto \cite{Tanaka88}
claimed that quantum Fisher information 
obtained from the success postselection alone could not overcome the QCRB.
However, there are still some benefits of using postselection measurements. 
There are reports on achieving the Heisenberg scaling 
of the single parameter estimation using postselection measurements 
\cite{Zhang114,Chen121,Pang113,Pang92,Jordan4,Jordan2}.
There may also be some advantages in suppressing certain types of technical noise
\cite{Jordan4,Pang92,Pang115,Harris118}, systematic errors \cite{Pang94}.
Especially, Jordan et al. claimed that for some special technical noise, indirect sensing gives higher 
Fisher information than a direct one \cite{Jordan4},
which was verified experimentally \cite{Viza92}.
The same advantage has been found when the correlation between success and failed postselections 
were taken into account \cite{Sinclair96}.

Recently, multiparameter estimations using postselection measurements have been attracting 
a lot of attention \cite{Vella122,Xia13,Jan116}.
However, the role of postselection measurements on
the saturation of simultaneous multiparameter 
estimation has not been fully discussed yet.

In this work, we discuss the multiparameter estimations 
under postselection measurements. We are interested in 
the general case when the sensor can be in mixed state. 
We take into account all the orthonormal postselected states,
(both the success and failed postselections.) 
Our work is significantly different from Tanaka and 
Yamamoto's work \cite{Tanaka88},
in which only a single parameter estimation in  
a success mode is considered.
We compare the total quantum Fisher information matrix (QFIM)
obtained from all the postselection measurements (named as $\bm{Q}$) 
and the general QFIM determined by the sensor state only ($\bm{H}$): 
$\bm{H}$ corresponds to the maximum information that the sensor has. 
Of course, the postselection measurements cannot allow us 
to obtain more information from the sensor than what it has. 
However,  as we show that the whole information 
in the sensor can be obtained via $\bm{Q}$ 
even if it is in mixed state while it cannot be done with direct measurements:
We found that $\bm{Q}$ can reach $\bm{H}$. 
We illustrate this framework in the estimation of
phase and its fluctuation since these topics have been
attracting a lot of attention recently
\cite{Vidrighin5,Altorio92,Szczykulska2,Sergey2013,Genoni106}.

This paper is organized as follows: Section~\ref{secii} 
introduces a measurement framework with postselection, and 
we formulate the Cram{\'e}r-Rao bounds, the condition for achieving these,  
and tradeoff relations. The application of our framework for measuring a phase and 
its fluctuation is presented in Sec.~\ref{seciii}.
We summarize the results and point out the benefit of 
measurements with postselections in Sec.~\ref{seciv}.
Appendixes provide supporting material.

\section{Estimation process with postselection}\label{secii}

A postselection measurement of a quantum system (called a sensor 
in this work) is as follows. 
It interacts with a so-called pointer which we 
call as a measurement apparatus (MA) in this work. 
After the interaction, it is postselected on 
a final state. Measuring the MA state reveals its information 
with the results of postselection. 

\subsection{Measurement process}\label{secii_subsecA}
We consider a quantum channel $\Lambda_{\bm \phi}$ that 
is characterized by  
a set of $d$ parameters  ($\bm{\phi} = \{\phi_1,\phi_2,...,\phi_d\}$)
to be estimated. We perform the following process.  
(\textbf{i}) 
A state $\rho_{\rm s,i}$ of the sensor is prepared. 
(\textbf{ii}) 
It evolves to $\rho'_{\rm s,i} = \Lambda_{\bm \phi} (\rho_{\rm s,i})$
after passing through the quantum channel $\Lambda_{\bm \phi}$,
which now contains the information of ${\bm \phi}$.
(\textbf{iii}) 
The sensor and MA  interact with each other and become a joint state  
\begin{align}\label{rho_sm}
 \rho_{\rm sm} &= \hat U_{\rm sm}\bigl(\rho'_{\rm s,i}
\otimes|\xi\rangle\langle\xi|\bigr)\hat U_{\rm sm}^\dagger, 
\end{align}
where  $|\xi\rangle$ is an initial MA state and 
\begin{align}\label{Usm}
 \hat U_{\rm sm} &= {\rm exp}(-ig\hat A_{\rm s}\otimes\hat M_{\rm m}),
\end{align}
is the unitary evolution caused by the sensor-MA interaction.  
$g$ is the interaction strength and can be controlled. 
$\hat A_{\rm s}$ and $\hat M_{\rm m}$ are operators on the sensor 
and the MA, respectively. 
The role of the interaction is to share the information of $\bm{\phi}$
in the sensor with the MA.
(\textbf{iv}) 
The sensor is postselected onto a final state 
$\rho_{\rm s,f}^{(\jmath)}
= |\psi_{\rm s,f}^{(\jmath)}\rangle\langle\psi_{\rm s,f}^{(\jmath)}|$. 
Here, we consider a set of orthonormal postselected states
with $\jmath = 1, 2, \dots$.

The probability of postselection into $\jmath$ is
\begin{align}\label{wc}
 w^{(\jmath)} &= {{\rm Tr}[(\rho_{\rm s,f}^{(\jmath)}
\otimes I_{\rm m})\rho_{\rm sm}]},
\end{align}
where $I_{\rm m}$ is the identity matrix in the MA space. 
When all orthonormal postselected states are
taking into account, we have
$\sum_\jmath w^{(\jmath)} = 1$.
The MA state after the postselection reads
\begin{align}\label{rho_pc}
\rho^{(\jmath)}_{\rm m} = \dfrac{{\rm Tr}_{\rm s}
[(\rho_{\rm s,f}^{(\jmath)}\otimes I_{\rm m})\rho_{\rm sm}]}{w^{(\jmath)}},
\end{align}
where ${\rm Tr}_{\rm s}[ * ]$ is a partial trace w.\ r.\ t.\ the sensor.
$\sum_\jmath \rho^{(\jmath)}_{\rm s,f} = I_{\rm s}$ 
by definition of the probability, where $I_{\rm s}$ is the identity 
matrix in the sensor space. Assume that we employ a POVM
$\hat\Pi_k$ on the final MA state 
for getting a measurement result $k$, 
the corresponding probability distributions are given as
\begin{align}\label{pro_k}
P(k|\jmath) = {\rm Tr_m}[\rho^{(\jmath)}_{\rm m}\hat\Pi_k].
\end{align}
$\bm{\phi}$ are estimated from these.

\subsection{Cram\'er-Rao bounds}\label{secii_subsecB}

Let us now define ${\bm F}$, a postselected classical 
Fisher information matrix (pCFIM). 
This is given by the probability distributions when measuring 
the final MA state in all the orthonormal postselected states
multiply by the corresponding probability \cite{Zhang114}.
Its elements are given as 
\begin{align}\label{CFI}
{\bm F}_{\alpha\beta} = 
\sum_\jmath w^{(\jmath)} \sum_{l}\dfrac{1}{P(l|\jmath)}
\dfrac{\partial P(l|\jmath)}{\partial \phi_\alpha}
\dfrac{\partial P(l|\jmath)}{\partial \phi_\beta}.
\end{align}
%
We also define a postselected quantum Fisher information matrix (pQFIM)
whose elements are
\begin{align}\label{QFI}
\bm{Q}_{\alpha\beta} = 
\sum_\jmath w^{(\jmath)} \ {\rm Tr_m} 
\Bigl[\rho^{(\jmath)}_{\rm m}
\dfrac{\hat{L}^{(\jmath)}_{\alpha}
\hat{L}^{(\jmath)}_{\beta}
+\hat{L}^{(\jmath)}_{\beta} 
\hat{L}^{(\jmath)}_\alpha}{2}\Bigr], 
\end{align}
%
where $\hat{L}_k^{(\jmath)}$ are 
symmetric logarithmic derivatives (SLDs) defined as  
\cite{Paris7,Braunstein72}
\begin{align}\label{SLD_def}
 \hat{L}_k^{(\jmath)} \rho^{(\jmath)}_{\rm m} 
+ \rho^{(\jmath)}_{\rm m} \hat{L}_k^{(\jmath)} 
&= 2 \frac{\partial \rho^{(\jmath)}_{\rm m}}{\partial\phi_k}.
\end{align} 
It is also worth mentioning that 
the general quantum Fisher information matrix (QFIM) of the sensor has elements 
\cite{Paris7,Helstrom1976}
\begin{align}\label{bQFI}
\bm{H}_{\alpha\beta} = {\rm Tr_s} 
\Bigl[\rho'_{\rm s,i}
\dfrac{\hat L^{(0)}_\alpha
\hat L^{(0)}_\beta+
\hat L^{(0)}_\beta\hat L^{(0)}_\alpha}{2}\Bigr],
\end{align}
where $\hat{L}^{(0)}_k$ are similarly defined 
with $\rho_{\rm s, i}'$ as Eq.~(\ref{SLD_def}).
The diagonal elements of $\bm{H}^{-1}$ provide the ultimate achievable precision 
of the estimations and are limited by quantum mechanics \cite{Braunstein72}.
The off-diagonal elements of $\bm{H}^{-1}$ provide the correlation between parameters.
In this work, we compare ${\bm{Q}}$ and $\bm{H}$ in term of a quantum tradeoff
as we will introduce below.

The precision of the estimation of $\bm\phi$ 
is evaluated by its covariance matrix $\bm{C}$ 
%
($\bm{C}_{\alpha\beta} = 
E[(\bm\phi-E[\bm\phi])(\bm\phi-E[\bm\phi])^T]$, 
where $E[ * ]$ is the expected value).
The diagonal element $\bm{C}_{\alpha\alpha}$ is 
the variance $(\delta\phi_\alpha)^2$.
We obtain the lower bounds for the covariance matrix as
\begin{align}\label{lower_bound}
M\bm{C}\ge {\bm F}^{-1}
\ge {\bm Q}^{-1}\ge {\bm H}^{-1},
\end{align}
where $M$ is the number of repeated measurements. 
See the proof in Appendix~\ref{appA}. 

In Eq.~\eqref{lower_bound}, the inequality
$M\bm{C}\ge {\bm F}^{-1}$ is the postselected
classical Cram\'er-Rao bound (pCCRB).
It may be saturated
by using a maximum likelihood estimator \cite{Braunstein25}. 
The saturation of pCCRB means that $M\bm{C} = {\bm F}^{-1}$.

The inequality
${\bm F}^{-1} \ge {\bm Q}^{-1}$ is referred to the postselected 
quantum Cram\'er-Rao bound (pQCRB). 
A POVM set that allows the saturation of the pQCRB (${\bm F} = {\bm Q}$)
is called optimal. Although the optimal POVMs
have been reported for single parameter estimations  
\cite{Vidrighin5,Crowley89,Dominik97}, 
it is now being actively studied for multiparameter estimations. 
The condition for 
${\bm F} = {\bm Q}$ 
is when $[\hat L_\alpha^{(\jmath)},
\hat L_\beta^{(\jmath)}] = 0$, 
or a weaker condition Tr$_{\rm m}[\rho_{\rm m}^{(\jmath)}
[\hat L_\alpha^{(\jmath)},
\hat L_\beta^{(\jmath)}]] = 0, \forall \jmath$, is satisfied
\cite{Szczykulska1,Gill61}.
%
We will discuss this condition
in Sec.~\ref{seciiC}.

Finally, the inequality ${\bm Q}^{-1}\ge {\bm H}^{-1}$ is due to the fact that 
maximum of ${\bm Q}$ depends on the choice of sensor and MA states.
We find that ${\bm Q} = {\bm H}$ can happen at a certain state of 
the sensor and the MA in our concrete example of a phase and its fluctuation 
measurement. See, Sec.~\ref{seciii}.  

\subsection{Condition for saturating the pQCRB}\label{seciiC}

So far, we note that the pCCRB 
[the first inequality in Eq. \eqref{lower_bound}] may be saturated 
by using a maximum likelihood estimator \cite{Braunstein25}, but 
the pQCRB [the second inequality in Eq. \eqref{lower_bound}]
is not easy to attain.
In principle, the pQCRB can be achieved if 
\begin{align}\label{eq:cond}
{\rm Tr} \bigl[\rho_{\rm m}
[\hat L_\alpha,
\hat L_\beta]] = 0,
\end{align} 
or much stronger condition of 
$[\hat L_\alpha,\hat L_\beta] = 0$,  
is satisfied \cite{Szczykulska1,Gill61}.
Here we omit the superscript $(\jmath)$ of 
$\rho_{\rm m}^{(\jmath)}, 
\hat L_\alpha^{(\jmath)}, 
\hat L_\beta^{(\jmath)}$ for short.
%
Further, condition~\eqref{eq:cond} can be expressed as 
\begin{align}\label{eq:cond_rho}
{\rm Im}\bigl[{\rm Tr} [\rho_{\rm m} L_\alpha L_\beta]\bigr]
= 0.
\end{align} 
Note that condition \refeq{eq:cond_rho} is generally valid 
even when a sensor is in mixed state~\cite{Szczykulska1,Rafal2020}. 

Matsumoto~\cite{Matsumoto_2002} proved that 
the QCRB can be achieved with a set of POVMs,
\begin{align}\label{eq:X}
\Pi_k &\equiv|X_k\rangle\langle X_k| , \\
|X_k\rangle &= \sum_{l}  [\bm Q]^{-1}_{k,l} \hat L_l|\psi\rangle,
\end{align}
in the case when a 
sensor is in pure state, 
i.e., $|\psi\rangle$.
It is worthy to note that such a choice of POVM
is not unique, for example, Humphreys et at. have introduced 
another method for choosing the POVMs at a fixed point
\cite{Humphreys111}.

The sensor is in mixed state in our case, and we should emphasize that it is not known how to construct a set of POVMs yet although the pQCRB should be achievable. 

\subsection{Tradeoff relations}\label{secii_subsecC}
From the inequality ${\bm F} \le {\bm H}$ in Eq.~\eqref{lower_bound},
we define a classical tradeoff ${\rm Tr}[{\bm F}{\bm H}^{-1}]$,
which quantifies how ${\bm F}$ can be close to ${\bm H}$.
This tradeoff is a kind of ``competition'' between the estimations of parameters. 
Similarly, the inequality ${\bm Q} \le {\bm H}$ leads to a tradeoff 
${\rm Tr}[{\bm Q}{\bm H}^{-1}]$,
which we refer as a quantum tradeoff.

In Sec.~\ref{seciii}, we will investigate these tradeoff relations 
in more concrete examples. 
We will show that 
the quantum tradeoff  ${\rm Tr}[{\bm Q}{\bm H}^{-1}]$ can reach 
the number of parameters with our proposed method, 
which implies that all the parameters 
are possible to attain the ultimate precision simultaneously.

\section{Simultaneous estimation of phase and fluctuation}
\label{seciii}

A phase fluctuation of a sensor and its phase may provide 
dynamical information of the environment surrounding the sensor. 
Therefore, the phase fluctuation can be a parameter of interest. 
There are several reports on the simultaneous phase and 
its fluctuation estimations, but they focused only on direct 
measurements
\cite{Vidrighin5,Altorio92,Szczykulska2,Sergey2013,Genoni106},
where the maximum classical tradeoff reached only one 
and could not reach two: Note that 
the number of parameters, i.e., $d = 2$
in Refs.~
\cite{Vidrighin5,Altorio92,Szczykulska2}.

We re-examine the quantum tradeoff relation in the case of 
postselection measurement with 
a discrete qubit MA. 
See Appendixes \ref{seciii_subsecA} and \ref{appB} 
for the case of a continuous Gaussian MA. 
We achieve 
${\rm Tr}[{\bm Q}{\bm H}^{-1}] = 2$ at a certain initial 
condition of the sensor and MA. It  implies  that 
we are able to extract all information from the sensor. 

\subsection{The QFIM ${\bm H}$}
We assume the initial sensor state is 
$\rho_{\rm s,i}=
\dfrac{1}{2}
\begin{pmatrix}
1 & 1 \\
1 & 1
\end{pmatrix}
$. 
After passing through the phase ($\phi$) and its fluctuation
($\Gamma$) channel, it evolves to \cite{Altorio92} 
\begin{align}
\label{rho(phi)}
\rho'_{\rm s,i} = 
\dfrac{1}{2}
\begin{pmatrix}
1 & e^{-i\phi -\Gamma^2}\\
e^{i\phi -\Gamma^2} & 1
\end{pmatrix}.
\end{align}
In this case, $\bm{\phi} =\{\phi,\Gamma\}$. 
The QFIM $\bm{H}$ related to this state is a diagonal matrix
and can be calculated from Eq.~\eqref{bQFI},
as follows 
\begin{align}
\label{QFIM_phi_Gam}
 \bm{H}=
\begin{pmatrix}
\bm{H}_{\phi\phi}    & \bm{H}_{\phi\Gamma}\\
\bm{H}_{\Gamma\phi} & \bm{H}_{\Gamma \Gamma}
\end{pmatrix}
= 
\begin{pmatrix}
e^{-2\Gamma^2}   & 0\\
0                & \dfrac{4\Gamma^2}{e^{2\Gamma^2}-1}
\end{pmatrix}. 
\end{align}
%
Note that $\hat{L}_\phi$ and $\hat{L}_\Gamma$ can be easily 
calculated according to Eq.~(\ref{SLD_def}) \cite{Paris7}.
Note that the classical tradeoff was reported to be less than  
two in Refs.~\cite{Vidrighin5,Altorio92}
in direct measurements.

\subsection{Measurement Scheme}\label{seciii_subsecB}

The initial state of a qubit MA \cite{Wu374} is 
$|\xi\rangle = \sin(\theta/2)|0\rangle + \cos(\theta/2)|1\rangle$,
while the postselected states are chosen to be  
$\rho_{\rm s,f}^{(\jmath)} 
= |\psi_{\rm s,f}^{(\jmath)}\rangle\langle\psi_{\rm s,f}^{(\jmath)}|$, 
for $\jmath = 1, 2$,
where $ |\psi_{\rm s,f}^{(1)}\rangle 
= \sin(\gamma/2)|0\rangle + \cos(\gamma/2)|1\rangle$
and its orthonormal state 
$|\psi_{\rm s,f}^{(2)}\rangle = |\psi_{\rm s,f}^{(1)\perp}\rangle
= \cos(\gamma/2)|0\rangle - \sin(\gamma/2)|1\rangle$.
%
We choose the evolution  of 
$\hat U_{\rm sm} = \exp(-ig\sigma_z\otimes |1\rangle\langle 1|)$
by the sensor-MA interaction which is a prototype one in 
modular-value-based measurements 
\cite{Ho383,Kedem105,Ho95,Ho59,Ho380,Cormann93}
and which is easy to realize 
\cite{Ho380,Cormann93}.
We take $g= \pi/2$. 


The probabilities 
$w^{(1)}$ and $w^{(2)}$ are calculated as 
\begin{align}\label{pro_qubit}
w^{(1)} &= \dfrac{1}{2}
\bigl(1 - e^{-\Gamma^2} \cos\theta\sin\gamma\cos\phi\bigr), \\
w^{(2)} &= 1-w^{(1)},
\end{align} 
according to Eq.~(\ref{wc}). 

According to the definition (Eq.~\eqref{rho_sm}), 
$\rho_{\rm sm}$ ($4 \times 4$ matrix) can be obtained. Then, 
$\rho^{(1)}_{\rm m}$ is given as,
\begin{align}\label{rhop_c_appC}
\rho^{(1)}_{\rm m}= 
\begin{pmatrix}
 \dfrac{\sin ^2 \frac{\theta }{2}\ (e^{\Gamma ^2}+\sin\gamma  \cos \phi )}
{e^{\Gamma ^2}- \cos\theta \sin\gamma \cos\phi } 
& -\dfrac{\sin \theta  (i e^{\Gamma ^2} \cos\gamma+\sin \gamma  \sin \phi ) }
{2(e^{\Gamma ^2}-  \cos \theta \sin\gamma \cos\phi )}\\
-\dfrac{\sin \theta  (-i e^{\Gamma ^2} \cos \gamma+\sin \gamma  \sin \phi  )}
{2(e^{\Gamma ^2}-\cos \theta \sin\gamma \cos\phi )} 
&\dfrac{ \cos ^2 \frac{\theta }{2} (e^{\Gamma ^2}-\sin \gamma  \cos \phi )}
{e^{\Gamma ^2}-\cos \theta \sin \gamma \cos \phi } \\
\end{pmatrix},
\end{align}
according to 
Eqs.~(\ref{rho_sm}, \ref{Usm}, \ref{rho_pc}). 
Then, 
\begin{align}
L^{(1)}_\phi &= 
\begin{pmatrix}
-\dfrac{2 \cos ^2 \frac{\theta }{2} \sin \gamma  \sin \phi }
{e^{\Gamma ^2}-\cos\theta \sin \gamma \cos \phi } 
& \dfrac{1}{\cot \theta -e^{\Gamma ^2}\csc \theta \csc \gamma \sec \phi } \\
\dfrac{1}{\cot \theta -e^{\Gamma ^2} \csc \theta \csc\gamma \sec\phi } 
& \dfrac{2  \sin ^2 \frac{\theta }{2} \sin\gamma \sin\phi}{e^{\Gamma ^2}
-\cos \theta \sin \gamma \cos \phi } \\
\end{pmatrix},
\label{La_appC}\\
L^{(1)}_\Gamma &= 
\dfrac{e^{\Gamma ^2} \Gamma  (\coth \Gamma ^2-1)}
{e^{\Gamma ^2}- \cos \theta \sin\gamma \cos\phi }\times
\begin{pmatrix}
 2 \cos ^2 \frac{\theta }{2} (1-e^{\Gamma ^2}  \sin\gamma \cos\phi) 
& \sin \theta (i \cos \gamma +e^{\Gamma ^2} \sin \gamma  \sin \phi) \\
\sin \theta (-i \cos \gamma+e^{\Gamma ^2} \sin \gamma \sin\phi )
& 2 \sin ^2 \frac{\theta }{2}(1+e^{\Gamma ^2} \sin \gamma\cos \phi ) \\
\end{pmatrix},
\label{Lb_appC}
\end{align}
are obtained. 
Similarly, we obtain
\begin{align}\label{rhop_x_appC}
\rho^{(2)}_{\rm m}= 
\begin{pmatrix}
\dfrac{\sin ^2 \frac{\theta }{2} (e^{\Gamma ^2}-\sin \gamma\cos \phi ) }
{e^{\Gamma ^2}+\cos \theta \sin \gamma  \cos \phi} & 
\dfrac{\sin \theta(ie^{\Gamma ^2}\cos \gamma +\sin\gamma  \sin \phi)}
{2(e^{\Gamma ^2}+\cos \theta \sin \gamma \cos \phi)} \\
 \dfrac{\sin \theta(-i e^{\Gamma ^2} \cos \gamma+\sin \gamma  \sin \phi) }
{2(e^{\Gamma ^2}+\cos \theta \sin \gamma \cos \phi)} 
& \dfrac{\cos ^2\frac{\theta}{2}(e^{\Gamma ^2}+\sin\gamma \cos\phi )}
{e^{\Gamma ^2}+ \cos \theta \sin\gamma \cos \phi} \\
\end{pmatrix},
\end{align}
\begin{align}
L^{(2)}_\phi &= 
\begin{pmatrix}
 \dfrac{2 \cos ^2 \frac{\theta }{2} \sin \gamma  \sin \phi }
{e^{\Gamma ^2}+ \cos \theta \sin \gamma \cos \phi } 
& \dfrac{1}{\cot \theta +e^{\Gamma ^2} \csc \theta \csc \gamma \sec \phi } \\
 \dfrac{1}{\cot \theta +e^{\Gamma ^2} \csc \theta \csc \gamma \sec \phi } 
& -\dfrac{2  \sin ^2 \frac{\theta }{2}\sin \gamma  \sin \phi }
{e^{\Gamma ^2}+\cos \theta \sin \gamma \cos \phi } \\
\end{pmatrix},
\label{La_x_appC}\\
L^{(2)}_\Gamma &= 
\frac{e^{\Gamma ^2} \Gamma  \left(\coth \Gamma ^2-1\right)}
{e^{\Gamma ^2} + \cos \theta \sin \gamma \cos \phi }\times
\begin{pmatrix}
 2 \cos ^2 \frac{\theta }{2} (1+ e^{\Gamma ^2} \sin\gamma \cos \phi) 
& - \sin \theta (i \cos \gamma +e^{\Gamma ^2}\sin \gamma \sin \phi) \\
 -\sin \theta(-i \cos \gamma + e^{\Gamma ^2}  \sin \gamma \sin \phi)  
& 2 \sin ^2 \frac{\theta }{2}(1-e^{\Gamma ^2}\sin \gamma \cos \phi )\\
\end{pmatrix}.
\label{Lb_x_appC}
\end{align}

Then, we obtain
\begin{align} 
{\rm Tr}_{\rm m}\bigl[\rho^{(1)}_{\rm m}[\hat{L}^{(1)}_\phi,
\hat{L}^{(1)}_\Gamma]\bigr] &= 
\frac{-4 i \Gamma e^{\Gamma^2}\sin^2 \theta \sin^2 \gamma \cos\gamma}
{(e^{\Gamma^2} -\cos\theta \sin\theta \cos\phi)^3}, \\
{\rm Tr}_{\rm m}\bigl[\rho^{(2)}_{\rm m}[\hat{L}^{(2)}_\phi,
\hat{L}^{(2)}_\Gamma]\bigr] &= 
\frac{4 i \Gamma e^{\Gamma^2} \sin^2 \theta \sin^2 \gamma \cos \gamma}
{(e^{\Gamma^2} +\cos \theta \sin \theta  \cos \phi )^3}.
 \label{trace_fail_mode}
\end{align}
Choosing $\gamma = \pi/2$, we have 
Tr$_{\rm m}[\rho^{(\jmath)}_{\rm m}[\hat{L}^{(\jmath)}_\phi,
\hat{L}^{(\jmath)}_\Gamma]] = 0$ for $\jmath = 1, 2$.

\subsection{pQFIM}
We calculate the pQFIM. 
By substituting $\rho_{\rm m}^{(\jmath)},L_\phi^{(\jmath)}$ and 
$L_\Gamma^{(\jmath)} $ at $\gamma = \pi/2$ into 
Eq.~\eqref{QFI}, we obtain $\bm Q$ as
\begin{align}
\bm Q_{\phi\phi} &= 
\dfrac{1}
{e^{2\Gamma^2}\csc^2\theta-\cot^2\theta}, \label{cFI_11} \\
\bm Q_{\Gamma \Gamma}&=
\dfrac{2\Gamma^2(1+\cot\Gamma^2)\sin^2\theta}
{e^{2\Gamma^2}-\cos^2\theta}
,\label{cFI_22}\\
\bm Q_{\phi \Gamma}&=\bm Q_{\Gamma \phi} =0. 
\end{align} 

\begin{figure} [t]
\centering
\includegraphics[width=8.6cm]{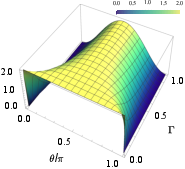}
\caption{
(Color online) Qubit MA: the quantum tradeoff ${\rm Tr}[{\bm Q}{\bm H}^{-1}]$ 
as a function of $\theta$ and $\Gamma$ is shown at $\gamma = \pi/2$. 
The tradeoff can reach its maximum (= 2 in this case) for 
a suitable choice of $\theta$, i.e., $\theta = \pi/2,$ regardless of 
$\Gamma$.
}
\label{fig2}
\end{figure}

The quantum tradeoff is given as 
\begin{align}\label{tradeoff_2_axam}
{\rm Tr}\bigl[{\bm Q}{\bm H}^{-1}\bigr] = 
\dfrac{{\bm Q}_{\phi\phi}}{{\bm H}_{\phi\phi}}+
\dfrac{{\bm Q}_{\Gamma \Gamma}}{{\bm H}_{\Gamma \Gamma}}=
\dfrac{2}{\csc^2\theta-e^{-2\Gamma^2}\cot^2\theta},
\end{align}
and shown as a function of $\theta$ and $\Gamma$  in Fig.~\ref{fig2}. 

For a suitable choice of MA state 
($\theta = \pi/2$), 
the quantum tradeoff ${\rm Tr}[{\bm Q}{\bm H}^{-1}]$ 
can reach two regardless of $\Gamma$.
This result implies that 
all the ratios ${\bm Q}_{\alpha\alpha}/{\bm H}_{\alpha\alpha},
\ \alpha = \phi, \Gamma,$ simultaneously reach one,
or ${\bm Q}$ approaches ${\bm H}$. Therefore, 
we should be possible to estimate both 
the phase and its fluctuation with the quantum-limit precision simultaneously. 

We note, however, that it is not possible to construct a set of 
POVMs corresponding to $\hat{L}_\phi^{(\jmath)}$ and 
$\hat{L}_\Gamma^{(\jmath)}$
because the sensor is in mixed state. See, Sec.~\ref{seciiC}. 
We believe that collective measurements,  
such as Refs.~\cite{Vidrighin5,Rafal2020,Ragy94}, 
may provide the way to measure $\phi$ and $\Gamma$ with ultimate 
precision simultaneously.

\section{Discussion and Conclusion}\label{seciv}
We analyze simultaneous multiple parameter estimations in postselection 
measurements in terms of tradeoff relations.  
We first derive classical and quantum Cram\'er-Rao lower bounds 
and discuss 
the tradeoffs in the postselection measurements in general. 
Then, we discuss simultaneous measurements of 
phase and its fluctuation with a discrete qubit MA. 
This example confirms our general results. 
We found that 
the tradeoffs can be saturated and thus 
all the parameters should, in principle, be possible to attain the ultimate precision 
simultaneously. 
We note, however, that it is not possible to construct a set of POVMs for 
such measurements because the sensor is in mixed state.
We have not yet been successful to propose a concrete 
measurement procedure and this is our future work.  

We conclude this work by pointing out that postselection measurements 
can provide another way to control a measurement  
through an extra freedom of postselected states and the MA state.

\begin{acknowledgements}
This work was supported by CREST(JPMJCR1774), JST.
\end{acknowledgements}

\appendix

\setcounter{equation}{0}
\renewcommand{\theequation}{A.\arabic{equation}}

\section{Proof of Cram{\'e}r-Rao bounds}
\label{appA}

We will prove Eq.~(\ref{lower_bound}) here. 

\subsection{Proof of $\bm F\le \bm Q$}

Let us recast $\bm F$ and $\bm Q$ as 
$\bm F = \sum_\jmath \bm{F}^{(\jmath)}$
and $\bm Q =  \sum_\jmath \bm{Q}^{(\jmath)}$.
Then, we can prove that $\bm F^{(\jmath)} \le \bm Q^{(\jmath)}$, 
$\forall \jmath$.

$\bm F^{(\jmath)} \le \bm Q^{(\jmath)}$ means 
$\bm u^\intercal[\bm F^{(\jmath)}]\bm u
\le 
\bm u^\intercal[\bm Q^{(\jmath)}]\bm u$ 
for arbitrary $d$-dimensional real vectors $\bm u$ \cite{Pezze119}. 
We first rewrite the elements of $\bm F^{(\jmath)}$ as 
${\bm F}^{(\jmath)}_{\alpha\beta}= 
w^{(\jmath)}\sum_\mu
[{\bm F}(\mu|\jmath)]_{\alpha\beta}$, 
where $[{\bm F}(\mu|\jmath)]_{\alpha\beta}$ is defined by  
\begin{align}\label{CFI_sup}
\notag [{\bm F}(\mu|\jmath)]_{\alpha\beta} &\equiv 
\dfrac{1}{P(\mu|\jmath)}
\dfrac{\partial P(\mu| \jmath)}{\partial \phi_\alpha}
\dfrac{\partial P(\mu|\jmath)}{\partial \phi_\beta}\\
&=\dfrac{1}{{\rm Tr_m}[\rho^{(\jmath)}_{\rm m}\hat\Pi_\mu]}
\dfrac{\partial {\rm Tr_m}[\rho^{(\jmath)}_{\rm m}
\hat\Pi_\mu]}{\partial \phi_\alpha}
\dfrac{\partial {\rm Tr_m}[\rho^{(\jmath)}_{\rm m}
\hat\Pi_\mu]}{\partial \phi_\beta},
\end{align}
where we have used $P(\mu|\jmath)={\rm Tr_m}
[\rho^{(\jmath)}_{\rm m}\hat\Pi_\mu]$.
Using the SLD, 
$\partial_k\rho^{(\jmath)}_{\rm m} = 
(\hat L^{(\jmath)}_k\rho^{(\jmath)}_{\rm m}
+\rho^{(\jmath)}_{\rm m}\hat L^{(\jmath)}_k)/2$, 
where $\partial_k\equiv \partial/\partial \phi_k$ 
and $k = \alpha, \beta$, we have
\begin{align}\label{Tr_P_sup}
\dfrac{\partial {\rm Tr_m}[\rho^{(\jmath)}_{\rm m}\hat\Pi_\mu]}
{\partial \phi_k}
&=
{\rm Tr_m}\Bigl[\dfrac{\partial 
\rho^{(\jmath)}_{\rm m}\hat\Pi_\mu}{\partial \phi_k}\Bigr]
\nonumber \\
&=
\dfrac{1}{2}\Bigl\{{\rm Tr_m}\bigl[\hat L^{(\jmath)}_k
\rho^{(\jmath)}_{\rm m}\hat\Pi_\mu\bigr]+
{\rm Tr_m}\bigl[\rho^{(\jmath)}_{\rm m}
\hat L^{(\jmath)}_k\hat\Pi_\mu\bigr]\Bigr\}
\nonumber \\
&=
{\rm Re}\Bigl[{\rm Tr_m}\bigl[\rho^{(\jmath)}_{\rm m}
\hat\Pi_\mu\hat L^{(\jmath)}_k\bigr]\Bigr]. 
\end{align}
%
%
The following equality is employed here. 
\begin{align}\label{cyclic_sup}
{\rm Tr_m}\bigl[\rho^{(\jmath)}_{\rm m}\hat L^{(\jmath)}_k\hat\Pi_\mu\bigr]=
{\rm Tr_m}\bigl[\hat L^{(\jmath)}_k\hat\Pi_\mu\rho^{(\jmath)}_{\rm m}\bigr]=
\Bigl[{\rm Tr_m}\bigl[\rho^{(\jmath)}_{\rm m}\hat\Pi_\mu
\hat L^{(\jmath)}_k\bigr]\Bigr]^*
\end{align}
%

We calculate as follows \cite{Yang2018}:
\begin{align}\label{F<Q_sup}
\notag 
\sum_{\alpha\beta} 
u_\alpha[\bm F(\mu|\jmath)]_{\alpha \beta}\ u_\beta &=
\dfrac{\Bigl[{\rm ReTr_m}\bigl[\rho^{(\jmath)}_{\rm m}
\hat\Pi_\mu\sum_\alpha u_\alpha \hat L^{(\jmath)}_\alpha\bigr]\Bigr]^2}
{{\rm Tr_m}[\rho^{(\jmath)}_{\rm m}\hat\Pi_\mu]}\\
\notag&\overset{(a)}{\le}\dfrac{\Bigl|{\rm Tr_m}
\bigl[\rho^{(\jmath)}_{\rm m}\hat\Pi_\mu
\sum_\alpha u_\alpha \hat L^{(\jmath)}_\alpha\bigr]\Bigr|^2}
{{\rm Tr_m}[\rho^{(\jmath)}_{\rm m}\hat\Pi_\mu]}\\
\notag&\overset{(b)}{\le}
\sum_{\alpha\beta}  u_\alpha u_\beta 
{\rm Tr_m} \bigl[\rho^{(\jmath)}_{\rm m}
\hat L^{(\jmath)}_\alpha\hat \Pi_\mu\hat L^{(\jmath)}_\beta\bigr]\\
\notag&\overset{(c)}{=}
\dfrac{1}{2}\sum_{\alpha\beta}u_\alpha 
{\rm Tr_m}\bigl[\rho^{(\jmath)}_{\rm m}
(\hat L^{(\jmath)}_\alpha\hat\Pi_\mu
\hat L^{(\jmath)}_\beta+\hat L^{(\jmath)}_\beta
\hat\Pi_\mu\hat L^{(\jmath)}_\alpha)\bigr] u_\beta\\
&\overset{(d)}{=}\sum_{\alpha\beta}u_\alpha 
[\bm Q(\mu|\jmath)]_{\alpha \beta}\ u_\beta. 
\end{align}
We employs the followings. 
(a) 
Using an inequality $[{\rm Re}(z)]^2 \le |z|^2$
for a complex number $z$.
(b) 
Using  the Cauchy-Schwartz inequality 
$|{\rm Tr}(\hat A^\dagger \hat B)|^2 \le 
{\rm Tr}(\hat A^\dagger \hat A){\rm Tr}(\hat B^\dagger \hat B)$, 
where $\hat A\equiv \sqrt{\hat\Pi_\mu}\sqrt{\rho_{\rm m}^{(\jmath)}}$ and 
$\hat B\equiv \sum_\alpha \sqrt{\hat\Pi_\mu} u_\alpha 
\hat L^{(\jmath)}_\alpha \sqrt{\rho_{\rm m}^{(\jmath)}}$. 
(c) 
Using the symmetry of the indices $\alpha$ and $\beta$. 
(d) Remembering the definition of  $\bm{Q}^{(\jmath)}$ 
($[\bm Q(\mu|\jmath)]_{\alpha\beta} \equiv 
{\rm Tr_m}\bigl[\rho^{(\jmath)}_{\rm m}
\frac{\hat L^{(\jmath)}_\alpha\hat\Pi_\mu
\hat L^{(\jmath)}_\beta+\hat L^{(\jmath)}_\beta
\hat\Pi_\mu\hat L^{(\jmath)}_\alpha}{2}\bigr]$). 
By taking the sum over $\mu$, 
we obtain 
\begin{align}\label{F<Q_sum_sup}
\sum_{\alpha\beta} u_\alpha[\bm F^{(\jmath)}]_{\alpha\beta}\ u_\beta \le 
\sum_{\alpha\beta} u_\alpha[\bm Q^{(\jmath)}]_{\alpha\beta}\ u_\beta.
\end{align}
Or, we obtain  $\bm F^{(\jmath)} \le \bm Q^{(\jmath)}$,
$\forall \jmath$,
and thus, we obtain $\bm F \le \bm Q$. $\Box$

\subsection{Proof of $\bm Q\le \bm H$}

We next prove the inequality $\bm Q\le \bm H$. 
Let us denote $\bm{Q}_{\rm sm}$ the Fisher information matrix
obtained from the joint measurements of the sensor-MA.
We note that $\bm{Q}_{\rm sm} = \bm{H}$ since no information
can be gained or lost after the sensor-MA interaction.
We, therefore, will prove that $\bm{Q} \le \bm{Q}_{\rm sm}$.
Remind that $\bm{Q} = \sum_\jmath \bm{Q}^{(\jmath)}$.
Assume that the existence of optimal measurement is
a set of POVMs $\{\Pi^{(\jmath)}_k\}$ for $k$ outcomes in the MA
and the corresponding probability is $P(k|\jmath)$, 
then we have 
\begin{align}\label{Qc_appA}
\bm Q^{(\jmath)}_{\alpha\beta} = 
w^{(\jmath)}
\sum_k\dfrac{1}{P(k|\jmath)}
\dfrac{\partial P(k|\jmath)}{\partial \phi_\alpha}
\dfrac{\partial P(k|\jmath)}{\partial \phi_\beta}.
\end{align}
%
%
%
The corresponding optimal POVM measurement of the joint state is given by a set
$\bigl\{\{\rho_{\rm s,f}^{(1)}\otimes\Pi^{(1)}_{k_1}\}, 
\{(\rho_{\rm s,f}^{(2)})\otimes\Pi^{(2)}_{k_2}\},
\{(\rho_{\rm s,f}^{(3)})\otimes\Pi^{(3)}_{k_3}\},...
\bigr\}$.
The probabilities are 
$w^{(\jmath)} P(k_i|\jmath)$, 
for $\jmath = 1, 2,...$.
The Fisher information corresponds to the joint sensor-MA state is
\begin{align}\label{Qsm_appA}
\bm Q_{\rm sm} &= 
\sum_\jmath \sum_{k_\jmath}\dfrac{1}{w^{(\jmath)} P(k_\jmath|{\jmath})}
\dfrac{\partial [w^{(\jmath)} P(k_\jmath|{\jmath})]}{\partial \phi_\alpha}
\dfrac{\partial [w^{(\jmath)} P(k_\jmath|{\jmath})]}{\partial \phi_\beta}
\nonumber \\
\notag&= 
\sum_\jmath \sum_{k_\jmath}\dfrac{1}{w^{(\jmath)} P(k_\jmath|{\jmath})}
\Bigl[\dfrac{\partial w^{(\jmath)}}{\partial\phi_\alpha}P(k_\jmath|\jmath)
+w^{(\jmath)}\dfrac{\partial P(k_\jmath|\jmath)}{\partial\phi_\alpha}\Bigr]
\Bigl[\dfrac{\partial w^{(\jmath)}}{\partial\phi_\beta}P(k_\jmath|\jmath)
+w^{(\jmath)}\dfrac{\partial P(k_\jmath|\jmath)}{\partial\phi_\beta}\Bigr]
\nonumber \\
\notag&=
\sum_\jmath w^{(\jmath)} \sum_{k_\jmath}\dfrac{1}{P(k_\jmath|{\jmath})}
\dfrac{\partial P(k_\jmath|{\jmath})}{\partial \phi_\alpha}
\dfrac{\partial P(k_\jmath|{\jmath})}{\partial \phi_\beta}
+
\mathcal{F}\\
\notag&= 
\sum_\jmath \bm{Q}^{(\jmath)} + \mathcal{F} \\
&\ge  \bm{Q}\;,
\end{align}
where $\mathcal{F}$ is given by
\begin{align}\label{F_appA}
\mathcal{F} = 
\sum_\jmath
\Bigl\{
\dfrac{1}{w^{(\jmath)}}
\dfrac{\partial w^{(\jmath)}}{\partial\phi_\alpha}
\dfrac{\partial w^{(\jmath)}}{\partial\phi_\beta}
+\sum_{k_\jmath}
\Bigl(\dfrac{\partial w^{(\jmath)}}{\partial\phi_\alpha}
\dfrac{\partial P(k_\jmath|{\jmath})}{\partial\phi_\beta}+
\dfrac{\partial w^{(\jmath)}}{\partial\phi_\beta}
\dfrac{\partial P(k_\jmath|{\jmath})}{\partial\phi_\alpha}\Bigr)
\Bigr\}.
\end{align}
Note that $\mathcal{F}$ is the classical Fisher information contributed by 
all the postselection probabilities, hence it is non-negative.  
As a result, we have $\bm Q \le \bm H. \ \Box$ 

\setcounter{equation}{0}
\renewcommand{\theequation}{B.\arabic{equation}}
\section{Continuous Gaussian MA}\label{seciii_subsecA}

In this Appendix, we provide another example
for the simultaneous estimation of phase and phase
fluctuation via a continuous measurement apparatus (MA). 
We consider the continuous Gaussian MA with a zero-mean 
in position $x$ 
\begin{align}
\notag
|\xi\rangle = 
\int dx \dfrac{1}{(2\pi\sigma^2)^{1/4}}
\exp\Bigl(-\dfrac{x^2}{4\sigma^2}\Bigr)|x\rangle, \nonumber 
\end{align}
where we take the natural unit so that $\hbar = 1$.
This MA is widely used in weak measurement studies 
\cite{Aharonov60,Ritchie66,Hosten319,
Dixon102,Brunner105,Zilberberg106,Gorodetski109}
and is a prototype for discussing the postselection measurements.
$|\xi\rangle$ is equivalently given by 
\begin{align}
\label{xi_Gauss}
|\xi\rangle 
= \int dp\Bigl(\dfrac{2\sigma^2}{\pi}\Bigr)^{1/4} \exp(-p^2\sigma^2) |p\rangle,
\end{align}
where $p$ is momentum. 

We consider the unitary evolution 
$\hat U_{\rm sm} = \exp(-ig\sigma_z\otimes \hat{p})$
as a sensor-MA interaction, where $\hat{p}$ is a momentum operator, 
$\hat{p} = -i \partial_x$.
Throughout this paper, we fix $g = \pi/2$ for simplicity. 
The postselected states are chosen 
the same as in the case of the qubit MA. 
Note that in this case we also have 
two orthonormal postselected states, i.e., $\jmath = 1, 2$.  
The probability of postselection, $w^{(\jmath)}$, 
is calculated according to 
Eq.~(\ref{wc}) where $I_{\rm m}$ is replaced with 
$\int_{-\infty}^\infty dp' |p'\rangle \langle p' |$. 
\begin{align}\label{Psp_Gauss}
\notag 
w^{(\jmath)} &= 
\Bigl(\dfrac{2\sigma}{\pi}\Bigr)^{\sfrac{1}{2}}\int dp \
e^{-2p^2\sigma^2} \
{\rm Tr}_{\rm s}
\bigl[\rho_{\rm s,f}^{(\jmath)} \hat U_{\rm s}\rho'_{\rm s,i}
\hat U^\dagger_{\rm s}\bigr] \\
&=\dfrac{1}{2}\Bigl(1+e^{-\Gamma^2-\frac{\pi^2}{8\sigma^2}}\cos\phi\sin\gamma\Bigr),
\end{align}
where 
$\hat U_{\rm s} = \exp(-i \frac{\pi}{2} p\sigma_z)$. 
Note that $p$ in 
$\hat{U}_{\rm s}$ is not an operator. 

We next decompose the sensor state as 
\begin{align}\label{rho'}
\rho'_{\rm s,i} = \sum_k\lambda_k
|\psi_k\rangle\langle\psi_k|
= \lambda_1|\psi_1\rangle\langle\psi_1| + 
\lambda_2|\psi_2\rangle\langle\psi_2|,
\end{align}
where $\lambda_k$ and $|\psi_k\rangle$ ($k = 1, 2$), are
eigenvalues and eigenstates of  $\rho'_{\rm s,i}$, respectively.
Substituting $\rho'_{\rm s,i}$ into Eq.~\eqref{rho_pc}, we have
\begin{align}\label{MAstate_Gauss}
\rho^{(\jmath)}_{\rm m} = \sum_{k=1}^2 \dfrac{\lambda_k
\hat B_k^{(\jmath)}|\xi\rangle\langle\xi
|[\hat{B}_k^{(\jmath)}]^\dagger}{w^{(\jmath)}}
= \sum_{k=1}^2\lambda_k|\xi^{(\jmath)}_k\rangle\langle\xi^{(\jmath)}_k|,
\end{align}
where 
$\hat B_k^{(\jmath)}= \bigl(\langle\psi_{\rm s,f}^{(\jmath)}|
\otimes\hat I_{\rm m}\bigr)
\hat U_{\rm sm}\bigl(|\psi_k\rangle\otimes\hat I_{\rm m}\bigr)$. 
We also define $|\xi^{(\jmath)}_k\rangle \equiv 
\hat B_k^{(\jmath)}|\xi\rangle/
\sqrt{w^{(\jmath)}}$.
We emphasize that, in general, $\{|\xi^{(\jmath)}_k\rangle\}$ are not orthogonal.
In the case of $\gamma = \pi/2$ and $\phi \rightarrow 0$, however, 
we have 
$\langle\xi^{(\jmath)}_1|\xi^{(\jmath)}_2 \rangle 
= \langle\xi^{(\jmath)}_2|\xi^{(\jmath)}_1 \rangle = 0$
(see Appendix \ref{appB}).
We define $N_k^{(\jmath)} \equiv \langle\xi^{(\jmath)}_k
|\xi^{(\jmath)}_k \rangle, k = 1, 2$.

We first show that 
Tr$_{\rm m}[\rho^{(\jmath)}_{\rm m}[\hat{L}^{(\jmath)}_\phi,
\hat{L}^{(\jmath)}_\Gamma]] $ is given as 
\begin{align}\label{condition_c} 
 {\rm Tr}_{\rm m}
\bigl[\rho^{(\jmath)}_{\rm m}[\hat{L}^{(\jmath)}_\phi,
\hat{L}^{(\jmath)}_\Gamma]\bigr] 
= 
4\sum_{k,l=1}^2\Bigl(\dfrac{\lambda_k}{N_l^{(\jmath)}}-
\dfrac{\lambda_l}{N_k^{(\jmath)}}\Bigr)
\dfrac{\langle\xi^{(\jmath)}_k|\partial_\phi\rho_{\rm m}^{(\jmath)}|
\xi^{(\jmath)}_l\rangle
\langle\xi^{(\jmath)}_l|\partial_\Gamma\rho_{\rm m}^{(\jmath)}
|\xi^{(\jmath)}_{k}\rangle
}
{(\lambda_kN_k^{(\jmath)}+\lambda_lN_l^{(\jmath)})^2},
\end{align} 
in the Gaussian MA case. Then 
${\rm Tr}_{\rm m}\bigl[\rho^{(\jmath)}_{\rm m}[\hat{L}^{(\jmath)}_\phi,
\hat{L}^{(\jmath)}_\Gamma]\bigr] = 0$ when  $\gamma=\pi/2$ and 
$\phi\to 0$. (See the detailed calculation in Appendix \ref{appB}.)
Using also collective measurements, for example, 
$\bm{Q}$ may be achieved or the pQCRB 
can satisfy.

We  analytically obtain $\bm{Q}$
with the conditions that $\gamma=\pi/2$ and $\phi\to 0$, as follows. 
\begin{align}\label{Q_Gauss}
\bm{Q}^{(\jmath)}_{\alpha\beta} = 
4 w^{(\jmath)} 
\sum_{k,l = 1}^2\dfrac{\lambda_k}{N_l^{(\jmath)}}
\dfrac{\langle\xi^{(\jmath)}_k|\partial_\alpha
\rho^{(\jmath)}_{\rm m}
|\xi^{(\jmath)}_l\rangle
\langle\xi^{(\jmath)}_l|\partial_\beta\rho^{(\jmath)}_{\rm m}
|\xi^{(\jmath)}_k\rangle}
{(\lambda_kN_k^{(\jmath)}+\lambda_lN_l^{(\jmath)})^2},
\end{align}
where we have defined $\bm{Q}^{(\jmath)}$,
$\jmath = 1, 2$ are the pQFIMs for
the two orthonormal postselected states, respectively.
See Appendix \ref{appB} for detailed calculations. 
Finally, we have the total pQFIM, 
$\bm{Q} = \bm{Q}^{(1)} + \bm{Q}^{(2)}$.
Straightforward calculations give $\bm{Q}$ as
\begin{align}
\bm{Q}_{\phi\phi} &= 
\dfrac{e^{-\Gamma^2}}
{\cosh\Gamma^2+\coth\bigl(\frac{\pi^2}{8\sigma^2}\bigr)
\sinh\Gamma^2}, \label{GMA_GG}\\
\bm{Q}_{\Gamma \Gamma} &=
2\Gamma^2 \ \text{csch}\Gamma^2\ 
\text{csch}\bigl(\Gamma^2+\frac{\pi^2}{8\sigma^2}\bigr)
\sinh\bigr(\frac{\pi^2}{8\sigma^2}\bigr), \label{GMA_pp}\\ 
\bm Q_{\phi \Gamma}&=\bm Q_{\Gamma \phi} =0. \label{GMA_pG}
\end{align}
where $\text{csch } x = 2/(e^x-e^{-x})$.

\begin{figure} [t]
\centering
\includegraphics[width=8.6cm]{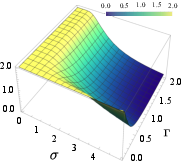}
\caption{
(Color online) Continuous Gaussian MA: 
the quantum tradeoff ${\rm Tr}[{\bm Q}{\bm H}^{-1}]$  
as a function of $\sigma$ and $\Gamma$ is shown at $\gamma = \pi/2$. 
The tradeoff can reach its maximum (= 2 in this case) for 
a suitable choice of $\sigma$, i.e., $\sigma < 1$ 
(weak measurement)  regardless of 
$\Gamma$.
}
\label{fig1}
\end{figure} 

We now calculate the quantum tradeoff.
By using Eqs.~(\ref{GMA_GG}, \ref{GMA_pp}) 
and Eq.~\eqref{QFIM_phi_Gam},
the quantum tradeoff, ${\rm Tr}[{\bm Q}{\bm H}^{-1}]$,  
in the simultaneous estimation of $\phi$ and $\Gamma$
reads
\begin{align}\label{Q_phi_Gamm_Gauss}
\notag
{\rm Tr}\bigl[{\bm Q}{\bm H}^{-1}\bigr] &=
\dfrac{{\bm Q}_{\phi\phi}}{{\bm H}_{\phi\phi}}+
\dfrac{{\bm Q}_{\Gamma \Gamma}}{{\bm H}_{\Gamma \Gamma}}\\
&=
2e^{\Gamma^2}\text{csch}\bigl(\Gamma^2+\frac{\pi^2}{8\sigma^2}\bigr)
\sinh\bigl(\frac{\pi^2}{8\sigma^2}\bigr).
\end{align}
The result is summarized in Fig. \ref{fig1}. 
The quantum tradeoff can reach the maximum of two 
for small $\sigma$ regardless of $\Gamma$,
which is consistent with 
${\rm Tr}_{\rm m}\bigl[\rho^{(\jmath)}_{\rm m}
[\hat{L}^{(\jmath)}_\phi,
\hat{L}^{(\jmath)}_\Gamma]\bigr] = 0$.
This result implies that 
all the ratios ${\bm Q}_{\alpha\alpha}/{\bm H}_{\alpha\alpha},
\ \alpha = \phi, \Gamma,$ simultaneously reach one.
It also means  that
it is not impossible to estimate both 
the phase and its fluctuation with the quantum-limit precision
simultaneously. 
This is the benefit of the postselection measurement scheme 
in which an extra freedom of MA, $\sigma$ in this case, 
is introduced. 
We also observe that ${\bm Q}_{\phi\phi}/{\bm H}_{\phi\phi} = 
{\bm Q}_{\Gamma \Gamma}/{\bm H}_{\Gamma \Gamma}$, or 
they can always attain the same precision: 
It is known as a Fisher-symmetric informationally complete (FSIC)
\cite{Li116}.

\setcounter{equation}{0}
\renewcommand{\theequation}{C.\arabic{equation}}
\section{Detailed Calculation in the case of  Continuous Gaussian MA}\label{appB}
\subsection{Calculation of ${\rm Tr}[\rho[L_\alpha,L_\beta]]$
in general}
Let us first calculate the SLD operator corresponding to 
an arbitrary $\rho$
 with the following Lyapunov representation \cite{Paris7}
\begin{align}\label{SLD_appB}
\hat L_\alpha = 2\int_0^\infty dt e^{-t\rho}
\bigl(\partial_\alpha\rho\bigr) e^{-t\rho},
\end{align} 
with 
$\rho = \sum_k\lambda_k|\xi_k\rangle\langle\xi_k|$,
where $\langle\xi_k|\xi_l\rangle = 0$ for $k\neq l$, and 
$\langle\xi_k|\xi_k\rangle = N_k$.
Evaluating the exponential $e^{-t\rho}$, we have
\begin{align}\label{exp_appB}
e^{-t\rho} = \sum_k \dfrac{e^{-t\lambda_kN_k}}{N_k}|\xi_k\rangle\langle\xi_k|.
\end{align} 
Substituting Eq.~\eqref{exp_appB} into Eq.~\eqref{SLD_appB}, we obtain 
\begin{eqnarray}\label{SLD_re_appB}
\hat L_\alpha = 2\sum_{kl}\dfrac{1}{N_kN_l}\ 
\dfrac{\langle\xi_k|\partial_\alpha\rho|\xi_l\rangle}
{\lambda_kN_k+\lambda_lN_l}
\
|\xi_k\rangle\langle\xi_l|.
\end{eqnarray} 
%
We next evaluate the term $\hat L_\alpha\hat L_\beta$ as
\begin{align}\label{LL_appB}
\hat L_\alpha\hat L_\beta = 4\sum_{kll'}\dfrac{1}{N_kN_lN_{l'}}\ 
\dfrac{\langle\xi_k|\partial_\alpha\rho|\xi_l\rangle}
{\lambda_kN_k+\lambda_lN_l}\
\dfrac{\langle\xi_l|\partial_\beta\rho|\xi_{l'}\rangle}
{\lambda_lN_l+\lambda_{l'}N_{l'}}\
|\xi_k\rangle\langle\xi_{l'}|.
\end{align} 
Using $\rho = \sum_n\lambda_n|\xi_n\rangle\langle\xi_n|$,
we have
\begin{align}\label{rhoLL_appB}
\rho\hat L_\alpha\hat L_\beta = 4\sum_{kll'}\dfrac{\lambda_k}{N_lN_{l'}}\ 
\dfrac{\langle\xi_k|\partial_\alpha\rho|\xi_l\rangle}
{\lambda_kN_k+\lambda_lN_l}\
\dfrac{\langle\xi_l|\partial_\beta\rho|\xi_{l'}\rangle}
{\lambda_lN_l+\lambda_{l'}N_{l'}}\
|\xi_k\rangle\langle\xi_{l'}|.
\end{align} 
Taking the trace, we obtain 
\begin{align}\label{Tr_rhoLL_appB}
{\rm Tr}[\rho\hat L_\alpha\hat L_\beta] = 
4\sum_{kl}\dfrac{\lambda_k}{N_l}\ 
\dfrac{\langle\xi_k|\partial_\alpha\rho|\xi_l\rangle
\langle\xi_l|\partial_\beta\rho|\xi_{k}\rangle
}
{(\lambda_kN_k+\lambda_lN_l)^2}.
\end{align} 
Similarly, we have
\begin{align}\label{Tr_rhoLbLa_appB}
{\rm Tr}[\rho\hat L_\beta\hat L_\alpha] = 
4\sum_{kl}\dfrac{\lambda_l}{N_k}\ 
\dfrac{\langle\xi_k|\partial_\alpha\rho|\xi_l\rangle
\langle\xi_l|\partial_\beta\rho|\xi_{k}\rangle
}
{(\lambda_kN_k+\lambda_lN_l)^2}.
\end{align} 
Finally, we obtain
\begin{align}\label{Cond_appB}
{\rm Tr}\bigl[\rho[\hat L_\alpha,\hat L_\beta]\bigr] = 
4\sum_{kl}\Bigl(\dfrac{\lambda_k}{N_l}-\dfrac{\lambda_l}{N_k}\Bigr)\ 
\dfrac{\langle\xi_k|\partial_\alpha\rho|\xi_l\rangle
\langle\xi_l|\partial_\beta\rho|\xi_{k}\rangle
}
{(\lambda_kN_k+\lambda_lN_l)^2},
\end{align} 
%
where
\begin{align}\label{partial_rho_appB}
\partial_\alpha\rho = 
\sum_n\Bigl(\partial_\alpha\lambda_n|\xi_n\rangle\langle\xi_n|
+\lambda_n|\partial_\alpha\xi_n\rangle\langle\xi_n|
+\lambda_n|\xi_n\rangle\langle\partial_\alpha\xi_n|\Bigr).
\end{align} 

\subsection{Calculation of ${\rm Tr}[\rho[L_\alpha,L_\beta]]$
in our example}\label{GMA_2}
Let us now apply the above calculations to our case of the 
continuous Gaussian MA
where ${\rm Tr}[\rho[L_\alpha,L_\beta]]$
now becomes ${\rm Tr}_{\rm m}[\rho_{\rm m}^{(\jmath)}
[L_\phi^{(\jmath)},L_\Gamma^{(\jmath)}]]$ for 
$\jmath = 1, 2$.
We start from the sensor state given in
Eq. \eqref{rho(phi)} and decompose it into the sum of the 
eigenstates as in  Eq.~\eqref{rho'}. 
We obtain 
\begin{align}\label{lamda_appB}
\lambda_1= \dfrac{1}{2}(1-e^{-\Gamma^2})\ \text{and } 
\lambda_2= \dfrac{1}{2}(e^{-\Gamma^2}+1),
\end{align}
\begin{align}\label{psi_k_appB}
|\psi_1\rangle= \dfrac{1}{\sqrt{2}}
\begin{pmatrix}-e^{-i\phi}\\1\end{pmatrix}
\ \text{and } 
|\psi_2\rangle=\dfrac{1}{\sqrt{2}}
\begin{pmatrix}e^{-i\phi}\\1\end{pmatrix}.
\end{align}
We next calculate $|\xi_k^{(\jmath)}\rangle$
which is defined by:
\begin{align}\label{xi_k^triangle_appB}
|\xi_k^{(\jmath)}\rangle&=\dfrac{\hat{B}_k^{(\jmath)}}
{\sqrt{w^{(\jmath)}}}|\xi\rangle,
\end{align}
where 
$\hat{B}_k^{(\jmath)} =  
\bigl(\langle\psi_{\rm s,f}^{(\jmath)}|\otimes\hat I_{\rm m}\bigr)
\hat U_{\rm sm}\bigl(|\psi_k\rangle\otimes\hat I_{\rm m}\bigr)$. 
%
We show explicitly
\begin{align}\label{B_kc_appB}
\notag\hat{B}_k^{(\jmath)} &=  
\Bigl(\langle\psi_{\rm s,f}^{(\jmath)}|
\otimes\int dp |p\rangle\langle p|\Bigr)
e^{-ig\sigma_z\otimes \hat p}
\Bigl(|\psi_k\rangle\otimes\int dp' |p'\rangle\langle p'\Bigr)\\
&=\int  dp\ \langle\psi_{\rm s,f}^{(\jmath)}|
e^{-igp\sigma_z}|\psi_k\rangle 
|p\rangle\langle p|.
\end{align}
Then we have
\begin{align}
\hat{B}_1^{(1)} &=  
\int  dp\ 
\Bigl[\dfrac{e^{-\frac{i(p\pi+2\phi)}{2}}\bigl(e^{i(p\pi+\phi)}
\cos\frac{\gamma}{2}-\sin\frac{\gamma}{2}\bigr)}
{\sqrt{2}}\Bigr]
|p\rangle\langle p|, \label{B_1c_appB} \\
\hat{B}_2^{(1)} &=
\int  dp\ 
\Bigl[\dfrac{e^{-\frac{i(p\pi+2\phi)}{2}}\bigl(e^{i(p\pi+\phi)}
\cos\frac{\gamma}{2}+\sin\frac{\gamma}{2}\bigr)}
{\sqrt{2}}\Bigr]
|p\rangle\langle p|, \\
\hat{B}_1^{(2)}&=  
\int  dp\ 
\Bigl[-\dfrac{e^{-\frac{i(p\pi+2\phi)}{2}}\bigl(
\cos\frac{\gamma}{2}+e^{i(p\pi+\phi)}\sin\frac{\gamma}{2}\bigr)}
{\sqrt{2}}\Bigr]
|p\rangle\langle p|, \\
\hat{B}_2^{(2)} &=  
\int  dp\ 
\Bigl[\dfrac{e^{-\frac{i(p\pi+2\phi)}{2}}\bigl(\cos\frac{\gamma}{2}
-e^{i(p\pi+\phi)}\sin\frac{\gamma}{2}\bigr)}
{\sqrt{2}}\Bigr]
|p\rangle\langle p|.\label{B_2x_appB}
\end{align}
Substituting into Eq. \eqref{xi_k^triangle_appB} we obtain
\begin{align}\label{xi_k^triangle_re_appB}
|\xi_k^{(\jmath)}\rangle=
\Bigl(\dfrac{2\sigma^2}{\pi}
\Bigr)^{1/4}
\dfrac{1}{\sqrt{w^{(\jmath)}}}
\int dp\ e^{-p^2\sigma^2}
\overline{B_k^{(\jmath)}}
|p\rangle,
\end{align}
where we have used
$\langle p|\xi\rangle = (2\sigma^2/\pi)^{1/4}\exp(-p^2\sigma^2)$
and $\overline{B_k^{(\jmath)}}$ are given by $[*]$ in 
Eqs. (\ref{B_1c_appB}-\ref{B_2x_appB}) above.

We calculate 
$\langle\xi_1^{(\jmath)}|\xi_2^{(\jmath)}\rangle$.
For $\phi\to 0$, we have
\begin{align}\label{condi_appB}
\langle\xi_1^{(1)}|\xi_2^{(1)}\rangle
= \dfrac
{e^{\Gamma^2+\frac{\pi^2}{8\sigma^2}}\cos\gamma}
{e^{\Gamma^2+\frac{\pi^2}{8\sigma^2}}+\sin\gamma},\ \text{and }
\langle\xi_1^{(2)}|\xi_2^{(2)}\rangle
= -\dfrac
{e^{\Gamma^2+\frac{\pi^2}{8\sigma^2}}\cos\gamma}
{e^{\Gamma^2+\frac{\pi^2}{8\sigma^2}}-\sin\gamma}.
\end{align}
Then, $|\xi_1^{(\jmath)}\rangle$ and $|\xi_2^{(\jmath)}\rangle$
are orthonormal when $\gamma=\pi/2$. We select 
$\gamma=\pi/2$ hereafter. 
We next calculate 
the normalized constants $N_k^{(\jmath)} = 
\langle\xi_k^{(\jmath)}|\xi_k^{(\jmath)}\rangle$ 
which are given as 
\begin{align}\label{N12_suc_appB}
N_1^{(1)} =1 - \dfrac{1 + e^{\Gamma^2}}{
 1 + e^{\Gamma^2 + \frac{\pi^2}{8 \sigma^2}}},\ \text{and } 
N_2^{(1)}=
1 + \dfrac{-1 + e^{\Gamma^2}}{
 1 + e^{\Gamma^2 + \frac{\pi^2}{8 \sigma^2}}},
 \end{align}
and 
\begin{align}\label{N12_fail_appB}
N_1^{(2)} =1 - \dfrac{1 + e^{\Gamma^2}}{
1 - e^{\Gamma^2 + \frac{\pi^2}{8 \sigma^2}}},\ \text{and } 
N_2^{(2)}=
1 + \dfrac{-1 + e^{\Gamma^2}}{
1 - e^{\Gamma^2 + \frac{\pi^2}{8 \sigma^2}}}.
 \end{align}
 
 Finally, we calculate Eq.~\eqref{Cond_appB} which is recast as
\begin{align}\label{Cond_re_appB}
{\rm Tr}_{\rm m}\bigl[\rho_{\rm m}^{(\jmath)}
[\hat L_\phi^{(\jmath)},\hat L_\Gamma^{(\jmath)}]\bigr] = 
4\sum_{k,l=1}^2\Bigl(\dfrac{\lambda_k}{N_l^{(\jmath)}}
-\dfrac{\lambda_l}{N_k^{(\jmath)}}\Bigr)\ 
\dfrac{\langle\xi_k^{(\jmath)}|\partial_\phi
\rho_{\rm m}^{(\jmath)}|\xi_l^{(\jmath)}\rangle
\langle\xi_l^{(\jmath)}|\partial_\Gamma
\rho_{\rm m}^{(\jmath)}|\xi_{k}^{(\jmath)}\rangle
}
{(\lambda_kN_k^{(\jmath)}+\lambda_lN_l^{(\jmath)})^2}.
\end{align} 
First we derive Eq.~\eqref{partial_rho_appB}:
\begin{align}\label{partial_rho_ex_appB}
\partial_\phi\rho_{\rm m}^{(\jmath)} = 
\sum_{n=1}^2\Bigl(\partial_\phi\lambda_n|
\xi_n^{(\jmath)}\rangle
\langle\xi_n^{(\jmath)}|
+\lambda_n|\partial_\phi\xi_n^{(\jmath)}\rangle\langle\xi_n^{(\jmath)}|
+\lambda_n|\xi_n^{(\jmath)}\rangle\langle\partial_\phi\xi_n^{(\jmath)}|\Bigr).
\end{align} 
Next we calculate the term 
$\langle\xi_k^{(\jmath)}|
\partial_\phi\rho_{\rm m}^{(\jmath)}
|\xi_l^{(\jmath)}\rangle$
in Eq. \eqref{Cond_re_appB}:
\begin{align}
\langle\xi_k^{(\jmath)}|
\partial_\phi\rho_{\rm m}^{(\jmath)}
|\xi_l^{(\jmath)}\rangle 
&= \sum_{n=1}^2\Bigl(
\langle\xi_k^{(\jmath)}|
\partial_\phi\lambda_n|\xi_n^{(\jmath)}\rangle\langle\xi_n^{(\jmath)}
|\xi_l^{(\jmath)}\rangle
+
\langle\xi_k^{(\jmath)}|
\lambda_n|\partial_\phi\xi_n^{(\jmath)}\rangle\langle\xi_n^{(\jmath)}
|\xi_l^{(\jmath)}\rangle
+
\langle\xi_k^{(\jmath)}|
\lambda_n|\xi_n^{(\jmath)}\rangle\langle\partial_\phi\xi_n^{(\jmath)}
|\xi_l^{(\jmath)}\rangle
\Bigr) \nonumber \\ 
&=\sum_{n=1}^2\Bigl(
\partial_\phi\lambda_n
\langle\xi_k^{(\jmath)}|
\xi_n^{(\jmath)}\rangle\langle\xi_n^{(\jmath)}
|\xi_l^{(\jmath)}\rangle
+\lambda_n
\langle\xi_k^{(\jmath)}|
\partial_\phi\xi_n^{(\jmath)}\rangle\langle\xi_n^{(\jmath)}
|\xi_l^{(\jmath)}\rangle
+\lambda_n
\langle\xi_k^{(\jmath)}|
\xi_n^{(\jmath)}\rangle\langle\partial_\phi\xi_n^{(\jmath)}
|\xi_l^{(\jmath)}\rangle
\Bigr)\nonumber \\
&=\sum_{n=1}^2\Bigl(
\partial_\phi\lambda_n
N_k^{(\jmath)}\delta_{k,n}
N_l^{(\jmath)}\delta_{l,n}
+\lambda_n
\langle\xi_k^{(\jmath)}|
\partial_\phi\xi_n^{(\jmath)}\rangle
N_l^{(\jmath)}\delta_{l,n}
+\lambda_n
N_k^{(\jmath)}\delta_{k,n}
\langle\partial_\phi\xi_n^{(\jmath)}
|\xi_l^{(\jmath)}\rangle
\Bigr). \nonumber 
\end{align}
Equation~\eqref{Cond_re_appB} is explicitly given

\begin{align}\label{Cond_re2_appB}
\notag &{\rm Tr}_{\rm m}\bigl[\rho_{\rm m}^{(\jmath)}
[\hat L_\phi^{(\jmath)},\hat L_\Gamma^{(\jmath)}]\bigr] \\
\notag &= 
4\Bigl(\dfrac{\lambda_1}{N_2^{(\jmath)}}
-\dfrac{\lambda_2}{N_1^{(\jmath)}}\Bigr)\times
\dfrac{\bigl(\lambda_1N^{(\jmath)}_1\langle
\partial_\phi\xi^{(\jmath)}_1|\xi^{(\jmath)}_2\rangle
+\lambda_2N^{(\jmath)}_2\langle\xi^{(\jmath)}_1
|\partial_\phi\xi^{(\jmath)}_2\rangle\bigr)
\bigl(\lambda_1N^{(\jmath)}_1\langle\xi^{(\jmath)}_2
|\partial_\Gamma\xi^{(\jmath)}_1\rangle
+\lambda_2N^{(\jmath)}_2\langle\partial_\Gamma\xi_2^{(\jmath)}
|\xi_1^{(\jmath)}\rangle\bigr)}
{(\lambda_1N_1^{(\jmath)}+\lambda_2N_2^{(\jmath)})^2}\\
&+4\Bigl(\dfrac{\lambda_2}{N_1^{(\jmath)}}
-\dfrac{\lambda_1}{N_2^{(\jmath)}}\Bigr)\times
\dfrac{\bigl(\lambda_1N^{(\jmath)}_1\langle\xi^{(\jmath)}_2|
\partial_\phi\xi^{(\jmath)}_1\rangle
+\lambda_2N^{(\jmath)}_2\langle\partial_\phi\xi^{(\jmath)}_2
|\xi^{(\jmath)}_1\rangle\bigr)
\bigl(\lambda_1N^{(\jmath)}_1\langle\partial_\Gamma\xi^{(\jmath)}_1
|\xi^{(\jmath)}_2\rangle
+\lambda_2N^{(\jmath)}_2\langle\xi_1^{(\jmath)}
|\partial_\Gamma\xi_2^{(\jmath)}\rangle\bigr)}
{(\lambda_1N_1^{(\jmath)}+\lambda_2N_2^{(\jmath)})^2}.\
\end{align} 
Indeed, we also calculate all inner products
$\langle\xi_m^{(\jmath)}|
\partial_\alpha\xi_n^{(\jmath)}\rangle
$
and their complex conjugations
similar as we did in Eqs. (\ref{condi_appB} - \ref{N12_fail_appB}).
We list them here:
\begin{align} 
\langle\xi_1^{(1)}|\partial_\phi\xi_1^{(1)}\rangle &=
\langle\partial_\phi\xi_1^{(1)}|\xi_1^{(1)}\rangle^\dagger=
-\frac{i}{2}\Bigl(1 - \dfrac{1 + e^{\Gamma^2}}{
1 + e^{\Gamma^2 + \frac{\pi^2}{8 \sigma^2}}}\Bigr)\\
\langle\xi_1^{(1)}|\partial_\phi\xi_2^{(1)}\rangle &=
\langle\partial_\phi\xi_2^{(1)}|\xi_1^{(1)}\rangle^\dagger=
\frac{i}{2}\Bigl(1 - \dfrac{1 + e^{\Gamma^2}}{
1 + e^{\Gamma^2 + \frac{\pi^2}{8 \sigma^2}}}\Bigr)\\
\langle\xi_2^{(1)}|\partial_\phi\xi_1^{(1)}\rangle &=
\langle\partial_\phi\xi_1^{(1)}|\xi_2^{(1)}\rangle^\dagger=
\frac{i}{2}\Bigl(1 + \dfrac{-1 + e^{\Gamma^2}}{
1 + e^{\Gamma^2 + \frac{\pi^2}{8 \sigma^2}}}\Bigr)\\
\langle\xi_2^{(1)}|\partial_\phi\xi_2^{(1)}\rangle &=
\langle\partial_\phi\xi_2^{(1)}|\xi_2^{(1)}\rangle^\dagger=
-\frac{i}{2}\Bigl(1 + \dfrac{-1 + e^{\Gamma^2}}{
1 + e^{\Gamma^2 + \frac{\pi^2}{8 \sigma^2}}}\Bigr)
\end{align}
\begin{align} 
\langle\xi_1^{(1)}|\partial_\Gamma\xi_1^{(1)}\rangle &=
\langle\partial_\Gamma\xi_1^{(1)}|\xi_1^{(1)}\rangle^\dagger=
\dfrac{-1+e^{\frac{\pi^2}{8 \sigma^2}}}
{\bigl(1+ e^{\Gamma^2 + \frac{\pi^2}{8 \sigma^2}}\bigr)^2}e^{\Gamma^2}\Gamma\\
\langle\xi_1^{(1)}|\partial_\Gamma\xi_2^{(1)}\rangle &=
\langle\partial_\Gamma\xi_2^{(1)}|\xi_1^{(1)}\rangle^\dagger=
0\\
\langle\xi_2^{(1)}|\partial_\Gamma\xi_1^{(1)}\rangle &=
\langle\partial_\Gamma\xi_1^{(1)}|\xi_2^{(1)}\rangle^\dagger=
0\\
\langle\xi_2^{(1)}|\partial_\Gamma\xi_2^{(1)}\rangle &=
\langle\partial_\Gamma\xi_2^{(1)}|\xi_2^{(1)}\rangle^\dagger=
\dfrac{1+e^{\frac{\pi^2}{8 \sigma^2}}}
{\bigl(1+ e^{\Gamma^2 + \frac{\pi^2}{8 \sigma^2}}\bigr)^2}e^{\Gamma^2}\Gamma
\end{align}
\begin{align} 
\langle\xi_1^{(2)}|\partial_\phi\xi_1^{(2)}\rangle &=
\langle\partial_\phi\xi_1^{(2)}|\xi_1^{(2)}\rangle^\dagger=
-\frac{i}{2}\Bigl(1 - \dfrac{1 + e^{\Gamma^2}}{
1 - e^{\Gamma^2 + \frac{\pi^2}{8 \sigma^2}}}\Bigr)\\
\langle\xi_1^{(2)}|\partial_\phi\xi_2^{(2)}\rangle &=
\langle\partial_\phi\xi_2^{(2)}|\xi_1^{(2)}\rangle^\dagger=
\frac{i}{2}\Bigl(1 - \dfrac{1 + e^{\Gamma^2}}{
1 - e^{\Gamma^2 + \frac{\pi^2}{8 \sigma^2}}}\Bigr)\\
\langle\xi_2^{(2)}|\partial_\phi\xi_1^{(2)}\rangle &=
\langle\partial_\phi\xi_1^{(2)}|\xi_2^{(2)}\rangle^\dagger=
\frac{i}{2}\Bigl(1 + \dfrac{-1 + e^{\Gamma^2}}{
1 - e^{\Gamma^2 + \frac{\pi^2}{8 \sigma^2}}}\Bigr)\\
\langle\xi_2^{(2)}|\partial_\phi\xi_2^{(2)}\rangle &=
\langle\partial_\phi\xi_2^{(2)}|\xi_2^{(2)}\rangle^\dagger=
-\frac{i}{2}\Bigl(1 + \dfrac{-1 + e^{\Gamma^2}}{
1 - e^{\Gamma^2 + \frac{\pi^2}{8 \sigma^2}}}\Bigr)
\end{align}
\begin{align} 
\langle\xi_1^{(2)}|\partial_\Gamma\xi_1^{(2)}\rangle &=
\langle\partial_\Gamma\xi_1^{(2)}|\xi_1^{(2)}\rangle^\dagger=
-\dfrac{1+e^{\frac{\pi^2}{8 \sigma^2}}}
{\bigl(1- e^{\Gamma^2 + \frac{\pi^2}{8 \sigma^2}}\bigr)^2}e^{\Gamma^2}\Gamma\\
\langle\xi_1^{(2)}|\partial_\Gamma\xi_2^{(2)}\rangle &=
\langle\partial_\Gamma\xi_2^{(2)}|\xi_1^{(2)}\rangle^\dagger=
0\\
\langle\xi_2^{(2)}|\partial_\Gamma\xi_1^{(2)}\rangle &=
\langle\partial_\Gamma\xi_1^{(2)}|\xi_2^{(2)}\rangle^\dagger=
0\\
\langle\xi_2^{(2)}|\partial_\Gamma\xi_2^{(2)}\rangle &=
\langle\partial_\Gamma\xi_2^{(2)}|\xi_2^{(2)}\rangle^\dagger=
\dfrac{1-e^{\frac{\pi^2}{8 \sigma^2}}}
{\bigl(1- e^{\Gamma^2 + \frac{\pi^2}{8 \sigma^2}}\bigr)^2}e^{\Gamma^2}\Gamma
\end{align}
Finally, by substituting all 
into Eq. \eqref{Cond_re2_appB} and doing some calculations,
we obtain 
\begin{align}\label{Cond_appB_ex_app}
{\rm Tr}_{\rm m}\bigl[\rho_{\rm m}^{(\jmath)}
[\hat L_\phi^{(\jmath)},\hat L_\Gamma^{(\jmath)}]\bigr] = 0.
\end{align} 

\subsection{The pQFIM}
Similarly as in Sec.~\ref{GMA_2}, we obtain 
\begin{align} \label{GMA_QD}
Q_{\alpha \beta}^{(\jmath)} 
= 4 w^{(\jmath)}
\sum_{k,l = 1}^2 \frac{\lambda_k}{N_l^{(\jmath)}}
\frac{\langle \xi_k^{(\jmath)} | \partial_\alpha \rho_{\rm m}^{(\jmath)} 
|\xi_l^{(\jmath)} \rangle \langle \xi_l^{(\jmath)} | \partial_\beta 
\rho_{\rm m}^{(\jmath)} |\xi_k^{(\jmath)} \rangle}
{(\lambda_k N_k^{(\jmath)} + \lambda_l N_l^{(\jmath)})^2}. 
\end{align}
By substituting $\langle \xi_k^{(\jmath)}| 
\partial_\alpha \rho_{\rm m}^{(\jmath)}| \xi_l^{(\jmath)} \rangle$'s 
calculated in Appendix~\ref{GMA_2} into Eq.~\eqref{GMA_QD}, 
we obtain $Q_{\alpha \beta}^{(\jmath)}$
and then Eqs.~\eqref{GMA_GG}, \eqref{GMA_pp} and \eqref{GMA_pG}. \\

\section*{AVAILABILITY OF DATA}
The data that support the findings of this study are available from the corresponding author upon reasonable request.



\end{document}